\documentclass[12pt]{iopart}
\usepackage{graphicx}
\begin{document}
\title[Shift of the orbit of Mercury and light bending by the Sun]{Complete calculations of the perihelion precession of Mercury and the deflection of light by the Sun in General Relativity}
\author{Christian Magnan}
\address{Coll\`ege de France, Paris; Universit\'e de Montpellier II-GRAAL, CNRS-UMR 5024, 34000 Montpellier, France}
\ead{magnan@graal.univ-montp2.fr}
\begin{abstract}
Taking up a method devised by Taylor and Wheeler and collecting pieces of their work we offer a self-contained derivation of the formulae giving both the precession of the orbit of a planet around the Sun and the deflection angle of a light pulse passing near the Sun in the framework of General Relativity. The demonstration uses only elementary algebra without resorting to tensor formalism. No prior knowledge in relativity is needed to follow the presentation.
\end{abstract}
\section{Introduction}
The deviation of light rays near the Sun and the change in the orientation of Mercury's orbit with each revolution were the first dramatic confirmations of the correctness of Einstein's theory of general relativity. Surprisingly it is uneasy to find the detailed calculations of those phenomena in the specialized scientific literature. For example even the comprehensive reference book \textit{Gravitation} by Misner \textit{et al} (1973) does not provide the reader with full explicit demonstrations of the effects. Truly it would seem that there is no simple way to conduct the complete calculation since it is necessary to dive into the full theory of general relativity. Fortunately, in their book about black holes and General Relativity, Taylor and Wheeler (2000) have succeeded in deriving the needed equations of motion without appealing to tensor formalism. To determine the motion of a free object in curved geometry, they use what they call a ``Principle of Extremal Aging'' along the path of the particle. That principle directly leads to the desired equations while using only simple algebra.

Exploiting their astute technique and collecting together parts of their work we carry out a complete presentation of the legendary effects predicted by Einstein when founding General Relativity, namely the deviation of stellar light near the sun's mass and the advance of the perihelion of Mercury. The aim of the present paper is to supply teachers, students and beginners in General Relativity with a document both yielding in a single place a full account of those relativistic phenomena and still remaining simple and accessible in spite of the difficulty of the problem. In fact in a sole article we begin with the derivation of the equations of motion in the curved spacetime surrounding a massive object and we end with the numerical evaluation of the magnitude of the physical effects under consideration. The calculation is thus entirely self-contained. Now, as far as we are aware, such a kind of document does not exist yet and thus deserves being published, even if it does not offer new material, as almost all the demonstrations shown here may be found in Taylor and Wheeler's book.

 We first closely follow Taylor and Wheeler's treatment in establishing the required formulae of motion in Schwarzschild geometry.  But we depart from their approach when calculating the advance of Mercury's perihelion. Carrying out a calculation adapted from Wald (1984), Taylor and Wheeler analyze the orbit of the planet by splitting its periodic motion into two components: an angular orbit and a radial oscillation. While the angular motion is described as nearly circular, the small inward and outward displacement is considered as a simple harmonic oscillator moving radially in a potential well. In this context it is found that the two corresponding periods differ by a tiny amount. As the oscillation takes place more slowly, it becomes late with respect to the period of revolution and this gives rise to the precession phenomenon. Contrary to this strategy we prefer to proceed more directly by adopting the algebraic method used by Weinberg (1972) in his book on General Relativity. The corresponding procedure allows the orbit to be elliptical and does not require the development of additional concepts. Actually, as they are written in polar coordinates, the equations of motion immediately yield the progression of the azimuthal angle $\phi$ as a function of the radial coordinate $r$ and thus make it possible to evaluate the change in this angle over one revolution. In particular the formulae show that from one passage at the perihelion to the next one the azimuthal angle $\phi$ increases by a little more than $2\pi$, which means that the perihelion does advance along the orbit. 
 
 Additionally, taking advantage of the expressions at hand we rediscover Kepler's third law relating the period of revolution of a planet to the length of its semi-major axis.
 
 We finally write the formulae determining the trajectory of a photon around a massive body and compute the change in the direction of a starlight pulse passing near the Sun. The integration of the equations readily shows that the azimuthal angle of the photon increases by more than $\pi$ along its path ($\pi$ corresponding to a straight line in the absence of mass), the additional quantity representing the relativistic deflection. 
 
 Incidentally, as the paper is intended to be fully accessible to students and (again) self-contained, some elementary notions of physics, especially in relativity, are recalled here and there. As a result no prerequisite in relativity is needed to study this article.

\section{Principle of the calculation}
The orbit of Mercury, the closest planet to the Sun, was known as early as 1859 to change with time in a partly unexplained way. It was observed that the perihelion (the orbital point closest to the Sun) moves a little around the Sun, advancing at a tiny rate with each orbit. But whereas Newtonian mechanics was able to account for the major part of this change, a residual of 43 seconds of arc per century remained unexplained. It is this residual that was correctly predicted by Einstein's theory.  Likewise Arthur Eddington's verification of the deflection of starlight near the Sun in the total solar eclipse of 29th May, 1919, gave Albert Einstein an immediate notoriety. Now while the qualitative aspect and the historical impact of both the precession of Mercury's perihelion and the deviation of light rays near the Sun are widely (and legitimately) described in the scientific literature as proofs of the rightness of General Relativity, the detailed calculations of the magnitude of those tiny relativistic effects are rarely given. The present paper is consecrated to this ``numerical exercise''.

General relativity teaches us that mass particles or photons propagating freely through space follow what is called a ``geodesic'' of spacetime. Thus we have to examine the following points:
\begin{itemize}
\item what is a ``geodesic''?
\item which equations govern a geodesic?
\item find the solutions of those equations for a mass particle and determine the orbit of a planet (which is actually an ellipse in a first approximation: thank you Newton!);
\item compute the tiny angle of rotation of the ellipse after the completion of one orbit;
\item by letting the mass of the particle tend towards zero, find the trajectory of light near the Sun; 
\item compute the angle between the initial and final directions of a light ray grazing the solar disk.
\end{itemize}

\section{Metrics for flat spacetime in special relativity}
Newtonian mechanics describes the motion of a particle in an absolute space with respect to an absolute time. The position of the moving object is located by its coordinates with respect to some frame of reference and is given as a function of time $t$. Special relativity affirms that there exists no absolute time and that time cannot be dissociated from space. The theory bases its reasoning on \textit{events}, each event being characterized by a point M (where it happens!) and a time $t$ (when it happens!). The events attached to a moving particle constitute what is called a \textit{worldline}.

Let us consider for example a spaceship moving freely through space, which means that all its motors are turned off. Let us imagine that regular flashes are emitted in accordance with a clock located inside the rocket and ticking at a constant rhythm. The time interval between two successive flashes will be denoted by $\tau$ (this quantity is thus measured with respect to the proper time of the spaceship). Think now of another frame of reference as constituted by an ensemble of space beacons, also free from acceleration, at constant mutual distances from one another (each free beacon stays at the same distance from its neighbours). Every signal bears the indication of its position in space (for instance by showing its distance from some point of origin) and holds its own clock. The clocks of this second frame are synchronized between them. Then in that latter frame the interval between two flashes (i.e. two events) is characterized by two numbers: the space interval $r$ and the time interval $t$. To determine those two quantities it suffices to record which beacon faces flash~\#1 and which beacon faces flash~\#2 while noting the times of those events as shown by the clocks attached to the beacons.

Special relativity is based on the following principle. The proper time interval $\tau$ between event~\#1 and event~\#2 is given by the formula
\begin{equation}
\tau^2 = t^2 - r^2
\end{equation}
and this quantity does not depend on the frame in which it is evaluated. In other words all observers agree on the value of $\tau$ computed by formula~(1), although the values of  $r$ and $t$ differ from one system of reference to another.

Unless otherwise indicated, distances will be measured in units of time, as is often done in astronomy. We have chosen to do so in writing (1). If however distances $r$ are expressed in conventional units, for instance in centimetres, then one should pass from the latter to our distance $r$ expressed in seconds via the formula $r \mbox{(in seconds)} = r \mbox{(in centimetres)}/c$  where $c$ is the speed of light in conventional units, namely $3\times10^{10}$~cm/s. (Expressing distances and times with the same unit would amount to taking the speed of light equal to unity.)

\section{Metrics for curved spacetime around a massive object}
In general relativity the property of invariance of the proper interval with respect to a change of the coordinates remains valid but only locally, i.e. under the condition of staying in a sufficiently small region of spacetime (its size depends on the accuracy of the measurements). The new feature concerns the expression of the proper time previously given by (1). The coefficients entering this formula depend now on the point of spacetime under consideration and the resulting expression takes the name of \textit{metrics}. In fact, and this is wonderful, the whole structure of spacetime, and especially its curvature, is included in the local expression of $\tau$ and in the form of its coefficients.

We are interested here in the structure of spacetime around the Sun. In order to describe the physics locally we consider two nearby events separated by infinitesimal amounts of time and space coordinates $dt$, $dx$, $dy$ and $dz$. If space were flat, the metrics would have the form
\begin{displaymath}(d\tau)^2 = (dt)^2 - (dx)^2 - (dy)^2 - (dz)^2 \end{displaymath}
which is usually written (and a little sloppily) by convention as
\begin{equation}d\tau^2 = dt^2 - dx^2 - dy^2 - dz^2 \,.\end{equation}
By working in spherical coordinates, in a plane containing the center of the Sun (this choice removes one spatial coordinate), that formula becomes
\begin{displaymath}d\tau^2 = dt^2 - dr^2 - r^2 d\phi^2 \,,\end{displaymath}
where $r$ denotes the distance to the center and  $\phi$ an azimuthal angle in the plane of the orbit.

But spacetime around a center of attraction of mass $M$ (for instance a black hole or the vicinity of the Sun) is not flat. It is characterized by the \textit{Schwarzschild metric} (see for instance Misner {\textit{et al}}, 1973)
\begin{equation}
d\tau^2 = (1 - 2M/r) dt^2 - (1 - 2M/r)^{-1} dr^2 - r^2 d\phi^2 \,.
\end{equation}

The story is really fantastic: the \textit{whole} structure of spacetime is embodied in this ``simple'' formula (3). Even the famous black hole lurks behind those apparently innocuous symbols.

One question: in which units is the mass $M$ expressed in that formula? It is seen that $M$ has the dimension of a length, a quantity that we measure here in seconds. Therefore $M$ will also be measured in seconds. The formula allowing to transform grams in seconds is
\begin{displaymath}M{\mbox{(in seconds})} =  (G/c^3) M{\mbox{(in grams)}}\end{displaymath}
where $(G/c^3) = 2,5\times10^{-39}\ \mbox{s/g}$. Thus the mass of the Sun is $2\times10^{33}$~g or $5\times10^{-6}$~s or 1.5~km.

\section{The equations of a geodesic}

The metric, that is the formula expressing at a given point of spacetime the temporal interval between two nearby events, reveals the presence of curvature as soon as the expression deviates from formula~(2) that corresponds to flat euclidian space. This metric will allow us to find the properties of the motion of a test particle free from acceleration. Actually both special and general relativity teach us that between two given events $E_1$ and $E_2$ a freely moving body follows the path for which the time interval $\tau$ is the maximum.  By defining a geodesic as the path maximising the time interval, it can also be stated that a freely moving particle follows a geodesic of spacetime.
\begin{quotation}
{\ }\\ 
{\textbf{Definition of a geodesic:}} \textit{the geodesic between two events $E_1$ and $E_2$ is the wordline for which the interval of proper time between $E_1$ and $E_2$ is maximum.}
\end{quotation}

That property of maximising the proper time will allow us to derive the equations of a geodesic. It will also yield the expressions of the energy and angular momentum of a particle in orbit around the centre of attraction.

\section{Energy of the particle}
Let us apply the principle of maximisation of the proper time interval in the following manner. Suppose that a free spatial ship (whose rockets are turned off) falls radially, therefore along a straight path, towards the central attractive mass. Imagine that three successive flashes, with nearby time and space coordinates, are emitted inside the spaceship. We observe those three events in some external frame. In that latter frame the event $E_1$ consists in the emission of a flash at time $t = 0$ when the spatial engine is located at radius $r_1$. The flash $E_2$ is emitted at time $t$ when the cabin is at radius $r_2$. The flash $E_3$ is emitted at time $T$ when the cabin is at radius $r_3$. The quantity $T$ is assumed to be small. We then imagine that we vary the intermediate coordinates of $E_2$. The principle of maximal aging says that the geodesic starting from $E_1$ and ending at $E_3$ will pass through event $E_2$ such that the proper time interval 
\begin{equation}
\tau = \tau_A + \tau_B,
\end{equation} 
is maximum. Here $\tau_A$ measures the interval over the first spacetime segment $A$, which connects $E_1$ to $E_2$ and $\tau_B$ measures the time interval over the second segment $B$, which connects $E_2$ to $E_3$.

In order to avoid varying all quantities at the same time, we assume in this experiment that the locations of the radii $r_1$, $r_2$ and $r_3$ are fixed and that only the time $t$, at which the second flash is emitted, is allowed to change. According to (3) the interval of proper time over the first segment $A$ is given by its square
\begin{equation}
{\tau_A}^2 = (1 - 2M/r_A) t^2 + \mbox{(terms without $t$)} \,,
\end{equation}
from which we deduce
\begin{equation}
\tau_A d\tau_A = (1 - 2M/r_A)tdt \,.
\end{equation}

The lapse of time over segment $B$ between the events $E_2$ and $E_3$ is $(T-t)$, and therefore the proper time duration $\tau_B$ is given by
\begin{equation}
{\tau_B}^2 = (1 - 2M/r) (T - t)^2 + \mbox{(terms without $t$)} \,,
\end{equation}
from which we deduce
\begin{equation}
\tau_B d\tau_B = - (1 - 2M/r_B)(T-t) dt \,.
\end{equation}

To make the total time interval $\tau = \tau_A + \tau_B$ maximum with respect to a variation $dt$ of the time $t$, we write
\begin{equation}
\frac{d\tau}{dt} = \frac{d\tau_A}{dt} + \frac{d\tau_B}{dt} = 0 \,.
\end{equation} 
Deducing $d\tau_A$ and $d\tau_B$ from (6) and (8) and letting quite naturally $t=t_A$ and $T-t = t_B$, we easily get
\begin{equation}
(1 -2M/r_A) (t_A/\tau_A) = (1 - 2M/r_B) (t_B/\tau_B)\ .
\end{equation}
The left side of that equation depends only on parameters characterizing the first segment A (which connects $E_1$ to $E_2$). The right side depends only on parameters related to the second segment B (which connects $E_2$ to $E_3$).

We have discovered in equation~(10) a quantity that is the same for both segment. This quantity is thus a constant of the motion for the free particle under consideration. For good physical reasons (especially to recover the formulae of special relativity), one is led to identify that constant of motion as the ratio of the energy of the particle to its mass. We write this very important result under the form
\begin{equation}
E/m = (1-2M/r) (dt/d\tau) \,,
\end{equation}
an expression in which we have returned to the differential notation for the intervals $t$ and $\tau$. 

Incidentally we may notice that with the units we have chosen, energy $E$ and mass $M$ are expressed in the same unit (for instance the second or the centimetre).

\section{Angular momentum of the particle}
We have applied the principle of maximising the proper time interval by varying the time of the intermediate event $E_2$. We now perform the same operation but this time we vary the angle $\phi$ of that intermediate event. We recall that $\phi$ measures the direction of the moving particle with respect to a direction chosen as the origin. We call it the \textit{azimuth}.

  We again consider the three events consisting in the emission of flashes inside a spaceship floating freely in space. The first segment $A$ connects event~$E_1$ to event~$E_2$. The second segment $B$ connects $E_2$ to $E_3$. The azimuthal angle of the first event is fixed at $\phi=0$. The angle of the last one is fixed at $\phi = \Phi$. The intermediate azimuth is taken as the variable $\phi$. Again in order not to vary everything at the same time, we assume that the radius $r$ at which the second flash is emitted stays constant.

We follow the same chain of reasoning as in the previous section. From the metric~(3), the time interval $\tau_A$ over the first segment is given by its square
\begin{equation}
{\tau_A}^2 = - {r_A}^2 \phi^2 + \mbox{(terms without $\phi$)} \,,
\end{equation}
and the interval $\tau_B$ over the second by
\begin{equation}
{\tau_B}^2 = - {r_B}^2 (\Phi - \phi)^2 + \mbox{(terms without $\phi$)} \,,
\end{equation}
from which we get
\begin{eqnarray}
\tau_A d\tau_A &=& -{r_A}^2\phi\, d\phi \\
\tau_B d\tau_B &=& {r_B}^2(\Phi - \phi) d\phi \,.
\end{eqnarray}
By writing $d\tau/d\phi = d(\tau_A + \tau_B)/d\phi = 0\ $, the equation
\begin{equation}
{r_A}^2 \phi_A/\tau_A = {r_B}^2 \phi_B/\tau_B 
\end{equation}
is easily obtained, similarly to (10), after having written quite naturally $\phi=\phi_A$ and $\Phi - \phi = \phi_B$. The left side, which contains only terms that are specific to the first segment, is equal to the right side, which contains only terms relative to the second segment. We thus exhibit another constant of motion, namely $r^2d\phi/d\tau$ (by shifting back to the differential notation), a quantity that turns out to be identified with the ratio of the angular momentum $L$ of the particle to its mass $m$, which we write as
\begin{equation}
L/m = r^2 (d\phi/d\tau) \,.
\end{equation}

\section{Computing the orbit}
Technically speaking in order to determine the trajectory of a moving body free from acceleration we apply the following strategy. Knowing the energy $E$ and the angular momentum $L$ of the particle of mass $m$
   ($E$ and $L$  depend on the initial conditions) we can follow the position of that particle by computing the increments of its spacetime coordinates $t$, $r$ and $\phi$  as the proper time $\tau$ itself advances. Algebraically for each increment $d\tau$ of the proper time we compute (or the computer calculates) the corresponding increments $dt$, $dr$ and $d\phi$ of the coordinate of the mobile body.  The squares of the increments $dt$ and $d\phi$ are extracted from (11) and (17) in the following form:
\begin{eqnarray}
dt^2 &=& (E/m)^2 (1-2M/r)^{-2} d\tau^2 \\
d\phi^2 &=& (L/m)^2 r^{-4} d\tau^2 \,.
\end{eqnarray}

We notice that the expression of $dr$ is missing. We get it by transporting the values of $dt$ and $d\phi$  into the metric equation (3) and solving it for $dr$. This yields
\begin{equation}
dr^2 = \left\{ (E/m)^2 - (1 - 2M/r)[1 + (L/m)^2r^{-2}]\right\} d\tau^2 \,.
\end{equation}

By dividing both sides of (20) and (19) we directly arrive to the equation of the orbit in polar coordinates as 
\begin{equation}
\left(\frac{1}{r^2} \frac{dr}{d\phi}\right)^2 = \left(\frac{E}{L}\right)^2 - \left(1 - \frac{2M}{r}\right) \left[\left(\frac{m}{L}\right)^2 + \frac{1}{r^2}\right]
\end{equation}

\section{Trajectory of the planet}
By making the change of variable
\begin{displaymath}u = 1/r, \ \ du=-dr/r^2\end{displaymath}
equation~(21) becomes
\begin{equation}
\left(\frac{du}{d\phi}\right)^2 = \frac{E^2}{L^2} - (1 -2Mu)\left(\frac{m^2}{L^2} + u^2\right) \,.
\end{equation}

We now follow the treatment by Weinberg (1972). Consider a test particle moving along its closed orbit around the Sun. Its distance $r$ to the central mass $M$ passes through a minimum $r_-$ and a maximum $r_+$, which correspond respectively to the perihelion and the aphelion. In accordance with the change of variable $u=(1/r)$ we let 
\begin{displaymath}u_-=1/r_-= v\ ,u_+=1/r_+=w \ ,\ \mbox{with}\ \  r_-<r<r_+\ , w < u < v\ .\end{displaymath}
At both points the derivative $dr/d\phi$ vanishes, which writes as:
\begin{eqnarray}
\frac{E^2}{L^2} - (1 - 2Mv)\left(\frac{m^2}{L^2} + v^2\right) &=& 0 \\
\frac{E^2}{L^2} - (1 - 2Mw)\left(\frac{m^2}{L^2} + w^2\right) &=& 0 \, .
\end{eqnarray}
It is easy to extract $E^2/L^2$ and $m^2/L^2$ from those equations as
\begin{eqnarray}
\frac{m^2}{L^2} = [v+w - 2M(v^2 +vw + w^2)]/2M \\
\frac{E^2}{L^2} = {(v+w)(1-2Mv)(1-2Mw)}/{2M} \,.
\end{eqnarray}
Expression~(22) thus takes the form
\begin{eqnarray}
\fl \left(\frac{du}{d\phi}\right)^2 = (v+w)(1-2Mv)(1-2Mw)/2M  
\\ - (1-2Mu)[v+w -2M(v^2 + vw + w^2) + 2Mu^2]/2M \,,\nonumber
\end{eqnarray}
which does vanish for $u=v$ and $u=w$.

Up to now we have not made any approximation but in order to pursue the calculation we will take advantage of the smallness of the term $Mu$ and ignore too small quantities. In fact the mass of the Sun is $5\times10^{-6}$~s (as was said above) or $1.5\times10^{5}$~cm and the radius of Mercury's orbit is about $6\times10^{12}$~cm or $2\times10^2$~s, which gives a dimensionless factor $Mu$ of order $2.5\times10^{-8}$. Multiplying both sides of (27) by the factor $(1 - 2Mu)^{-1}$ we feel secure to develop that term up to the second order in M as
\begin{displaymath}(1 - 2Mu)^{-1} \simeq 1 + 2Mu + 4M^2u^2 \ .\end{displaymath}
We thus get
\begin{eqnarray}
\fl 
(1 - 2Mu)^{-1} \left(\frac{du}{d\phi}\right)^2 = (v+w)(1-2Mv)(1-2Mw)(1+2Mu+4M^2u^2)/2M \nonumber\\
-[v+w -2M(v^2 + vw + w^2) + 2Mu^2]/2M \,.
\end{eqnarray}

The trick to simplify the apparently quite complicated right side consists in noticing that we are dealing with a quadratic function of $u$ which vanishes at $u=v$ and $u=w$ (if we neglect the terms of order $M^3$). Therefore it has the form 
\begin{displaymath}C(v-u)(u-w)\,.\end{displaymath} The value of the constant $C$ is immediately obtained by letting $u=0$ as
\begin{displaymath}C = 1 - 2M (v+w) \,,\end{displaymath}
which allows us to write equation~(28) for the trajectory in the following quite compact form
\begin{equation}
\left(\frac{du}{d\phi}\right)^2 = (1-2Mu)[1 -2M(v+w)](v-u)(u-w)\ .
\end{equation}

By taking the square root of both sides of that equation and by neglecting terms of order $M^2$ or higher the trajectory of the particle around the Sun can be computed by integrating the expression
\begin{equation}
\frac{d\phi}{du}= \pm\frac{1 + M(u+v+w)}{\sqrt{(v-u)(u-w)}} \ .
\end{equation}

The integration is trivial if one makes the change of variable $u \rightarrow\psi$ defined by
\begin{equation}
u = \frac{1}{2} (v+w) + \frac{1}{2} (v-w)\cos\psi
\end{equation}
which easily leads to
\begin{displaymath}\sqrt{(v-u)(u-w)} = \frac{1}{2} (v-w) \sin\psi = - du/d\psi\end{displaymath}
or
\begin{equation}
\frac{du}{\sqrt{(v-u)(u-w)}} = -d\psi \,.
\end{equation}
We count the angles starting at the perihelion. At that point $\phi=\psi=0$ and $u=v$. From that point $r$   increases and thus $u=1/r$ decreases while the angle $\psi$ increases. Taking into account (30), (31) and (32) we arrive at the very simple integral
\begin{equation}
\phi(u) = \int_0^{\psi(u)} \left[1 + \frac{3M}{2}(v+w) + \frac{M}{2}(v-w)\cos\psi\right]\,d\psi \,.
\end{equation}

The first term inside the brackets, which is equal to unity, leads to the classical Newtonian ellipse. Actually if $\phi=\psi$, expression~(31) yields the equation of the trajectory in polar coordinates $(r,\, \phi)$ in the form
\begin{equation}
u \equiv \frac{1}{r} = \frac{1}{2} (v+w) + \frac{1}{2} (v-w)\cos\phi \,,
\end{equation}
which does correspond to an ellipse. Ordinarily the equation is written as
\begin{equation}
r = \frac{p}{1 + e \cos\phi} 
\end{equation}
where $p$ is the ellipse parameter (sometimes called the {\it semi-latus rectum}) and $e$ its eccentricity.

Identifying (34) and (35) we see that
\begin{equation}
\frac{1}{p} = \frac{1}{2} (v+w) = \frac{1}{2}\left(\frac{1}{r_-} + \frac{1}{r_+}\right)
\end{equation}
and
\begin{equation}
e = \frac{v - w}{v + w} = \frac{r_+ - r_-}{r_+ + r_-} \ .
\end{equation}
With common notations the major axis of the ellipse is 
\begin{equation}
2a = r_- + r_+ 
\end{equation}
and the following relation holds
\begin{displaymath}p=a\,(1- e^2) \ .\end{displaymath}
\section{Advance of the perihelion}
The second term in integral (33), namely $(3M/2)(v+w)$, reveals that the angle $\phi$ will increase more than twice $\pi$ when the planet returns to its perihelion, from which it started. This phenomenon precisely represents the precession of the orbit. In fact we multiply by 2 the increase in the angle $\phi$ between the perihelion $r_-$ (corresponding to $\psi=0$) and the aphelion $r_+$ (corresponding to $\psi=\pi)$ to obtain a shift per period equal to
\begin{equation}
\Delta\phi = \frac{6M\pi}{p}
\end{equation}
where $p$ is defined in (36). In conventional units we have
\begin{displaymath}
\Delta\phi = \frac{6\pi G}{c^2} \frac{M(\mbox{g})}{p\,(\mbox{cm})}\ .
\end{displaymath}
\begin{figure}
\begin{center}\includegraphics*[scale=0.8]{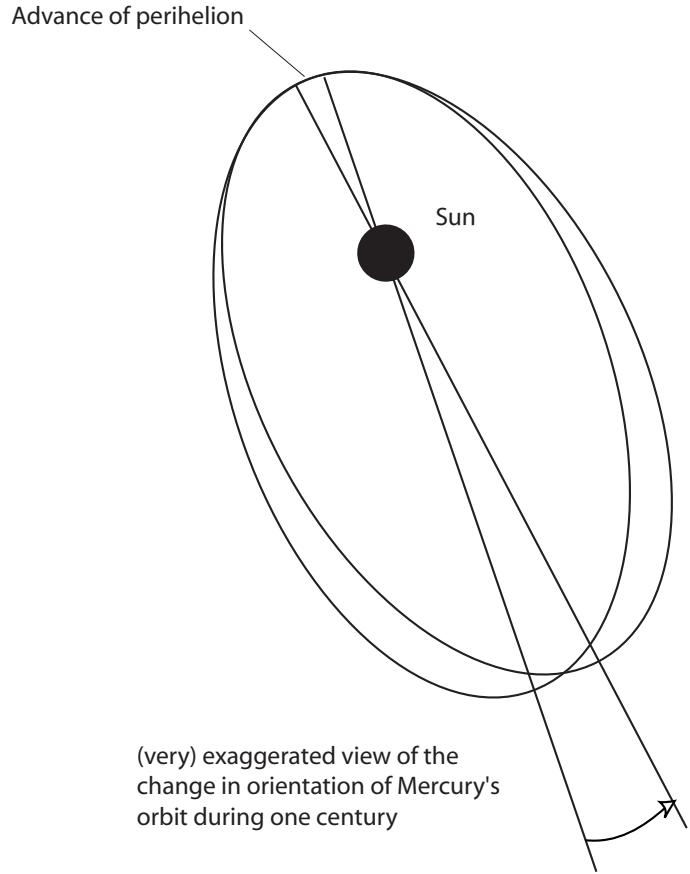}\end{center}
\caption{Observation shows that the perihelion of Mercury's orbit advances a little with each revolution. This means that the major axis of the ellipse rotates at a tiny rate. General relativity accounts for the magnitude of that effect, which is called the \textit{precession of the perihelion}.}
\end{figure}

The third sinusoidal term in $\sin\psi$ in integral (33) adds only a periodical perturbation which produces a kind of noise.

Let us calculate the numerical value of the angle of precession. The semimajor axis of the elliptical orbit of Mercure is $a=5.8\times10^{12}\ \mbox{cm}$ and its eccentricity is $e=0.206$. Thus $p=5.55\times10^{12}\ \mbox{cm}$. The mass of the Sun in centimetres is $M(\mbox{cm})=(G/c^2)M(\mbox{g})$, which yields (with $M=2\times10^{33}$~g) 
 $M=1.5\times10^5$~cm. Thus
\begin{displaymath}
\Delta\phi=5\times10^{-7}\ \mbox{radian/revolution} = 0.103"\mbox{/revolution}
\end{displaymath}
 
Knowing that there are 415 revolutions per century we conclude that the advance of the perihelion amounts to
\begin{displaymath}
\Delta\phi = 43"\  \mbox{per century}.
\end{displaymath}
 
\section{Period of revolution}
Let us profit from those formulae as a chance to calculate the period of revolution $T$ of a planet around the Sun and recover the Newtonian result. Formula~(17) teaches us that the increment $(1/2)r^2 d\phi=(1/2)(L/m)d\tau$ is proportional to the time increment $d\tau$. But that quantity is nothing else than the elementary area $dA$ swept out by the radius joining the Sun to the planet. Therefore the total area swept out at time $\tau$  since time $\tau=0$ is
\begin{equation}
A= (1/2)(L/m)\tau
\end{equation}
After one complete revolution the total time elapsed is $T$ and the total swept out area equals the area of the ellipse, namely $\pi ab$ if $a$ and $b$ denote the semiaxes. Thus
\begin{equation}
\pi ab = (1/2)(L/m)T
\end{equation}
or
\begin{equation}
T= 2\pi (m/L)\, a b \,.
\end{equation}
Taking into account (25) which gives (ignoring the $M$ term inside the parentheses)
\begin{displaymath}\frac{m^2}{L^2}=\frac{1}{2}(v+w)/M = \frac{1}{pM} \end{displaymath}
and known relations
\begin{displaymath}p = a(1-e^2)\ ,\ b=a\sqrt{1-e^2}\ ,\end{displaymath}
we find
\begin{equation}T^2 =\frac{4\pi^2 a^3}{M} \ .\end{equation}
In conventional units the formula reads 
\begin{displaymath}T_{\mbox{\small{s}}}^2 /a_{\mbox{\small{cm}}}^3 = 4\pi^2/(G M_{\mbox{\small{g}}}) \ ,\end{displaymath}
an expression in which the velocity of light does not appear. This is third Kepler's law.

\section{Trajectory of light}
The preceding treatment is apparently not relevant to the case of a photon. In fact the calculation of the trajectory was done by incrementing the proper time but this latter concept has no meaning for a photon since the interval between two events that are located on the wordline of a photon is always equal to zero (since at light velocity $r = t$, the interval  $\tau^2 = t^2 - r^2$ vanishes). 

Nevertheless it can be shown rigorously (see for instance Taylor and Wheeler, 2000) that by letting the mass of the particle tend towards zero, one arrives at the right results. Thus for $m=0$ our equation (21) takes the form
\begin{equation}
\left(\frac{1}{r^2} \frac{dr}{d\phi}\right)^2 = \left(\frac{E}{L}\right)^2 - \left(1 - \frac{2M}{r}\right) \frac{1}{r^2}
\end{equation}
That equation will allow us to determine the deviation of light rays passing near the Sun. 

It is necessary to specify the parameters found in the formulae. First the angular momentum of the moving particle at infinity is equal by definition to the product of its linear momentum  $p$ by what is called the \textit{impact parameter} $b$, which represents the distance between the center of attraction (the Sun in the present case) and the initial direction of the velocity of the particle (see the figure).

\begin{figure}
\begin{center}\includegraphics*[scale=0.9]{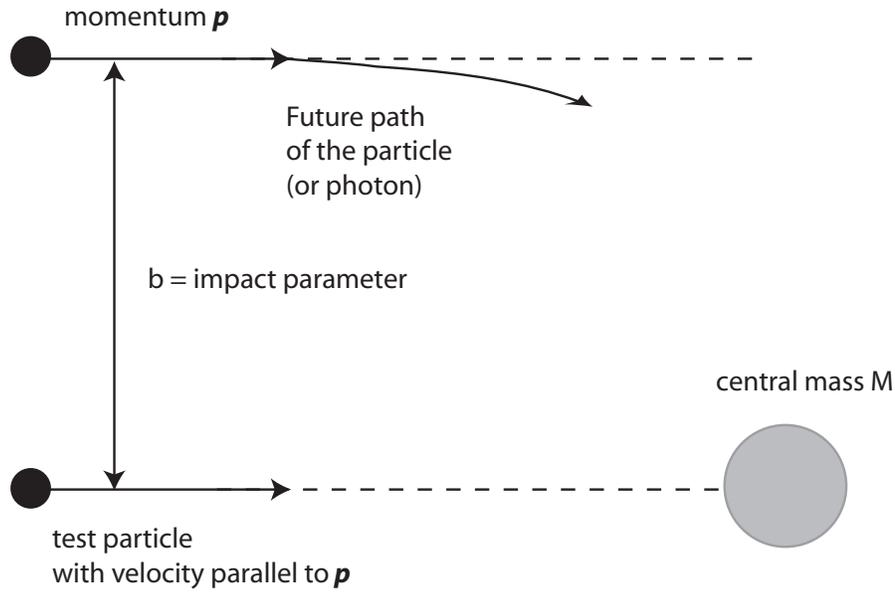}\end{center}
\caption{Definition of the impact parameter $b$. The moving particle approaches the mass $M$ from a great distance with vector momentum $\mathbf{p}$. A test particle with a parallel velocity plunges radially onto the mass $M$. The distance $b$ between their initially parallel paths at "infinity" is the impact parameter $b$.}
\end{figure}

 In other words
\begin{equation}
L = p\ b \,.
\end{equation}
In addition it is known that the momentum $p$ of a photon is equal to its energy $E$ (with the units that were chosen). It results at once from this formula that 
\begin{equation}
L/E = b \ .
\end{equation}
If the ratio $L/E$ is equal to the impact parameter $b$, (44) writes as
\begin{equation}
\left(\frac{1}{r^2} \frac{dr}{d\phi}\right)^2 = \frac{1}{b^2} - \left(1 - \frac{2M}{r}\right) \frac{1}{r^2} \ .
\end{equation}

\section{Determination of the deflection angle}
Formula (47) will allow us to determine the change in the direction of a light pulse caused by the gravitational field of the Sun. To achieve this aim we have to sum up the successive infinitesimal increments $d\phi$ of the azimuthal angle $\phi$ along the path. This means that we have to carry out the integration of $(d\phi/dr)dr$ when $r$ varies from the minimum distance denoted $R$ ($R$ is the radius of the Sun if the light ray grazes its surface) to infinity. We should still multiply that quantity by 2 to account for both symmetrical ``legs'' of the trajectory (the photon first approaches the Sun then recedes from it).

It is necessary to stipulate a further point, namely the relation existing between the two quantities $b$ and $R$ that we have introduced, and that are not independent. The point $r=R$ corresponds to the place where the light photon is closest to the Sun. There the photon moves tangentially. Since at that point there is no radial component, we can write that the derivative $dr/dt$ vanishes. It suffices to take the element $dr$ from (47) to immediately find 
\begin{equation}
\frac{1}{b^2} = \left(1 - \frac{2M}{R}\right) \frac{1}{R^2} \ ,
\end{equation}
so that this same equation (47) becomes 
\begin{equation}
\left(\frac{1}{r^2} \frac{dr}{d\phi}\right)^2 =  \left(1 - \frac{2M}{R}\right)\frac{1}{R^2} -\left(1 - \frac{2M}{r}\right) \frac{1}{r^2} \ .
\end{equation}

The form of the expression dictates to us to pose
\begin{displaymath}u = R/r\end{displaymath}
where $u$ varies between 1 and 0. The last equation (49) then becomes
\begin{eqnarray}
(du/d\phi)^2 &=& (1 - 2M/R) - (1 - 2Mu/R) u^2 \nonumber \\
&\mbox{or}& \nonumber \\
(du/d\phi)^2 &=& 1 - u^2 -(2M/R)(1 - u^3) \ .
\end{eqnarray}

Consequently the infinitesimal variation $d\phi$ of the azimuth is given in terms of the variation $du$ of $R/r$ by
\begin{eqnarray}
d\phi &=& [1 - u^2 - (2M/R)(1 - u^3)]^{-1/2}\ du \nonumber \\
     &=& \frac{(1-u^2)^{-1/2} du }{[1 - (2M/R)(1-u^3)(1 - u^2)^{-1}]^{1/2}} \ .
\end{eqnarray}

The presence of the term $(1 - u^2)$ in (51) encourages us to make the change of variable
\begin{displaymath}u = \cos \alpha , \ 0<u<1, \ 0<\alpha<\pi/2\end{displaymath}
which leads to
\begin{equation}
d\phi = \left[1 - (2M/R)(1 - \cos^3\alpha)\sin^{-2}\alpha \right]^{-1/2} d\alpha \,.
\end{equation}

By observing that
\begin{displaymath}\frac{1 - \cos^3\alpha}{\sin^2\alpha} = \frac{(1 -\cos\alpha)(1 + \cos\alpha +\cos^2\alpha)}{(1-\cos\alpha)(1+\cos\alpha)} = \cos\alpha + \frac{1}{1 + \cos\alpha}\end{displaymath}
we end up with the final equation of the trajectory under the form
\begin{equation}
d\phi=\left[1 - (2M/R)\left( \cos\alpha + \frac{1}{1+\cos\alpha}\right)\right]^{-1/2}d\alpha
\end{equation}
with
\begin{displaymath}\cos\alpha = R/r \,.\end{displaymath}

It is interesting to emphasize that so far there have been no approximation. This is quite rewarding.

\section{Approximations and integration}
The small value of the term $M/R$ will allow us to make an approximation and this operation will make us able to complete the integration. In conventional units the mass of the Sun is $2\times10^{33}$~grams and its radius is  $7\times10^{10}$~centimetres. By using the factor $G/c^2 = 7,4\times10^{-29}$~cm/g which makes it possible to transform grams into centimetres, we get the dimensionless quantity 
\begin{displaymath}M/R = (G/c^2) M(\mbox{in grams})/R(\mbox{in centimetres}) = 2 \times 10^{-6} \,.\end{displaymath}

In (53) we can thus use the classical approximation $(1 + \epsilon)^p \simeq 1 + p\,\epsilon$ to arrive at
\begin{equation}
d\phi =  \left[1 + (M/R)\left( \cos\alpha + \frac{1}{1+\cos\alpha}\right)\right] d\alpha \,.
\end{equation}
Therefore the total variation of the azimuth $\phi$ along the path of the photon is
\begin{eqnarray}
\phi &=& 2 \int_0^{\pi/2} \left[1 + (M/R)\left( \cos\alpha + \frac{1}{1+\cos\alpha}\right)\right] d\alpha \\
&=& 2 \left[ \alpha + \frac{M}{R}\left(\sin\alpha + \tan\frac{\alpha}{2}\right) \right]_0^{\pi/2}\\
&=&\pi + 4M/R
\end{eqnarray}
\begin{figure}
\begin{center}\includegraphics*[scale=0.7]{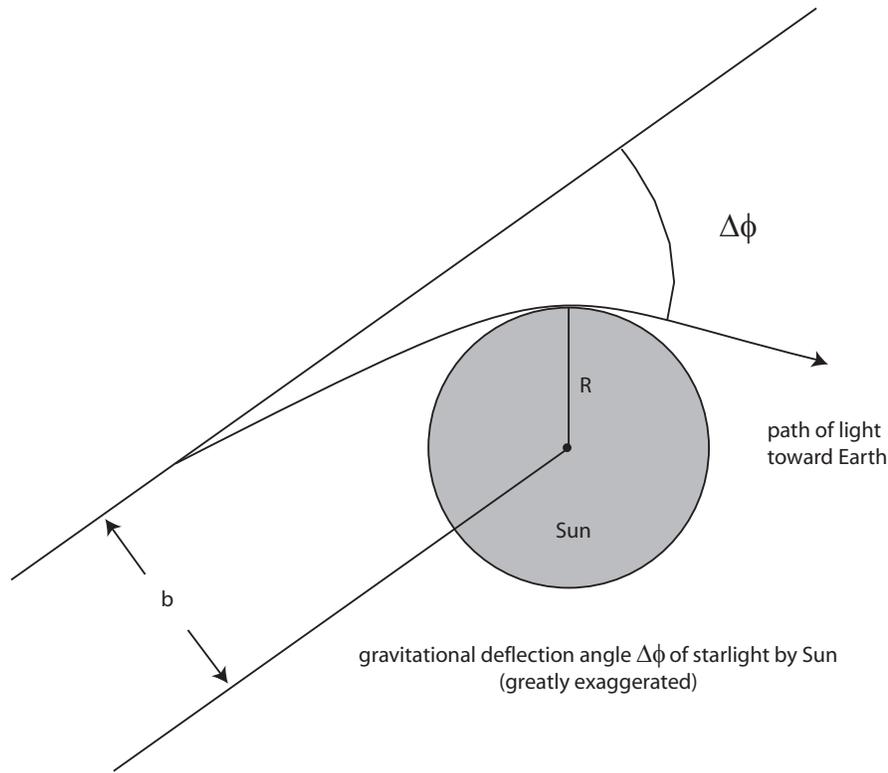}\end{center}
\caption{If there were no deflection the azimuthal angle $\phi$ would vary from 0 to $\pi$. In the presence of a mass the angle changes by an additional amount $\Delta\phi$.}
\end{figure}

The first term $\pi$ gives the total change in the azimuthal angle of the photon where there is no Sun present, since in that case the photon follows a straight path. It is the second term that gives the additional angle of deflection $\Delta\phi$ with respect to this straight line  
\begin{eqnarray}
\Delta\phi &=& 4 M/R  \\
\mbox{or in conventional units}& & \nonumber\\
\Delta\phi &=& 4 (G/c^2) M(\mbox{grams})/R(\mbox{centimetres}) \,.
\end{eqnarray}

Numerically at the surface of the Sun (with the values of the mass and the radius given above) one finds $\Delta\phi = 8,5 \times 10^{-6}$~radian, or ( knowing that $\pi$ radians equal 180 degrees and that there are 60 minutes of arc in one degree and 60 seconds of arc in one minute of arc)
\begin{displaymath}\Delta\phi = 1.75" \,.\end{displaymath}
\section{Summary}
The deflection of light rays by the Sun and the advance of the perihelion of Mercury were among the first observational confirmations of General Relativity and are thus of major historical value. Unfortunately, in spite of their importance, it is difficult for students or non-specialists to access to the full calculations of those effects as they are not readily found in the textbooks or in the scientific literature dedicated to Einstein's theory of gravitation. Using a method devised by Taylor and Wheeler (2000) and repeating sections of their book, the present work responds to this lack in providing an entirely self-contained derivation of the equations and complete algebraic calculations of Mercury's precession and starlight deflection. Moreover the paper uses only elementary algebra as it does not recourse to tensor formalism.
\section{Acknowledgment}
I wish to thank Christophe Mercier for his help in producing the figures, Lesley Hine for her careful revision of my english and Ana Palacios for her wise advice to publish among the astrophysical community the two pages devoted to the subject already available on the web (Magnan, 2006).
\section*{References}
\begin{harvard}
\item[] Magnan C 2006 http://www.lacosmo.com/DeflectionOfLight/index.html\\
http://www.lacosmo.com/PrecessionOfMercury/index.html
\item[] Misner C W, Thorne K S and Wheeler J A 1973 \textit{Gravitation} (San Francisco: Freeman)
\item[] Taylor E F and Wheeler J A 2000 \textit{Exploring Black Holes: Introduction to General Relativity} (Addison Wesley Longman)
\item[] Wald R M 1984 \textit{General relativity} (University of Chicago Press) pp 142-3
\item[] Weinberg S 1972 \textit{Gravitation and cosmology: principles and applications of the General Theory of Relativity} (John Wiley \& Sons) pp 194-8
\end{harvard}
\end{document}